\documentclass[11pt]{article}\textwidth 6.5in\textheight 9in
\usepackage{amssymb}\usepackage[colorlinks]{hyperref}\usepackage{color}
	\usepackage{stmaryrd}\usepackage{mathrsfs}
	\usepackage{graphicx}
	\usepackage{amsmath}
\usepackage{caption}
\usepackage[numbers]{natbib}

\begin{document}
\setlength{\captionmargin}{27pt}
\newcommand\hreff[1]{\href {http://#1} {\small http://#1}}
\newcommand\trm[1]{{\bf\em #1}} \newcommand\emm[1]{{\ensuremath{#1}}}
\newcommand\prf{\paragraph{Proof.}}\newcommand\qed{\hfill\emm\blacksquare}

\newtheorem{thr}{Theorem} 
\newtheorem{lmm}{Lemma}
\newtheorem{cor}{Corollary}
\newtheorem{con}{Conjecture} 
\newtheorem{prp}{Proposition}

\newtheorem{blk}{Block}
\newtheorem{dff}{Definition}
\newtheorem{asm}{Assumption}
\newtheorem{rmk}{Remark}
\newtheorem{clm}{Claim}
\newtheorem{example}{Example}

\newcommand{\ab}{a\!b}
\newcommand{\yx}{y\!x}
\newcommand{\yux}{y\!\underline{x}}

\newcommand\floor[1]{{\lfloor#1\rfloor}}\newcommand\ceil[1]{{\lceil#1\rceil}}

\newcommand{\lea}{<^+}
\newcommand{\gea}{>^+}
\newcommand{\eqa}{=^+}

\newcommand{\lel}{<^{\log}}
\newcommand{\gel}{>^{\log}}
\newcommand{\eql}{=^{\log}}

\newcommand{\lem}{\stackrel{\ast}{<}}
\newcommand{\gem}{\stackrel{\ast}{>}}
\newcommand{\eqm}{\stackrel{\ast}{=}}

\newcommand\edf{{\,\stackrel{\mbox{\tiny def}}=\,}}
\newcommand\edl{{\,\stackrel{\mbox{\tiny def}}\leq\,}}
\newcommand\then{\Rightarrow}

\newcommand\km{{\mathbf {km}}}\renewcommand\t{{\mathbf {t}}}
\newcommand\KM{{\mathbf {KM}}}\newcommand\m{{\mathbf {m}}}
\newcommand\md{{\mathbf {m}_{\mathbf{d}}}}\newcommand\mT{{\mathbf {m}_{\mathbf{T}}}}
\newcommand\K{{\mathbf K}} \newcommand\I{{\mathbf I}}

\newcommand\II{\hat{\mathbf I}}
\newcommand\Kd{{\mathbf{Kd}}} \newcommand\KT{{\mathbf{KT}}} 
\renewcommand\d{{\mathbf d}} 
\newcommand\D{{\mathbf D}}

\newcommand\w{{\mathbf w}}
\newcommand\Ks{\mathbf{Ks}} \newcommand\q{{\mathbf q}}
\newcommand\E{{\mathbf E}} \newcommand\St{{\mathbf S}}
\newcommand\M{{\mathbf M}}\newcommand\Q{{\mathbf Q}}
\newcommand\ch{{\mathcal H}} \renewcommand\l{\tau}
\newcommand\tb{{\mathbf t}} \renewcommand\L{{\mathbf L}}
\newcommand\bb{{\mathbf {bb}}}\newcommand\Km{{\mathbf {Km}}}
\renewcommand\q{{\mathbf q}}\newcommand\J{{\mathbf J}}
\newcommand\z{\mathbf{z}}

\newcommand\B{\mathbf{bb}}\newcommand\f{\mathbf{f}}
\newcommand\hd{\mathbf{0'}} \newcommand\T{{\mathbf T}}
\newcommand\R{\mathbb{R}}\renewcommand\Q{\mathbb{Q}}
\newcommand\N{\mathbb{N}}\newcommand\BT{\{0,1\}}
\newcommand\FS{\BT^*}\newcommand\IS{\BT^\infty}
\newcommand\FIS{\BT^{*\infty}}\newcommand\C{\mathcal{L}}
\renewcommand\S{\mathcal{C}}\newcommand\ST{\mathcal{S}}
\newcommand\UM{\nu_0}\newcommand\EN{\mathcal{W}}

\newcommand{\supp}{\mathrm{Supp}}

\newcommand\lenum{\lbrack\!\lbrack}
\newcommand\renum{\rbrack\!\rbrack}
\renewcommand\i{\mathbf{i}}
\renewcommand\qed{\hfill\emm\square}

\title{\vspace*{-3pc} The Kolmogorov Birthday Paradox}

\author {Samuel Epstein\footnote{JP Theory Group. samepst@jptheorygroup.org}}

\maketitle

\renewcommand{\contentsname}{\centering Contents}
\begin{abstract}
We prove a Kolmogorov complexity variant of the birthday paradox. Sufficiently large random subsets of strings are guaranteed to have two members $x$ and $y$ with low $\K(x/y)$. To prove this, we first show that the minimum conditional Kolmogorov complexity between members of finite sets is very low if they are not exotic. Exotic sets have high mutual information with the halting sequence.
\end{abstract}

\section{Introduction}
We prove a Kolmogorov complexity version of the birthday paradox. If you randomly select $2^{n/2}$ strings of length $n$, then, with overwhelming probability, you will have selected at least two strings $x$ and $y$ with low $\K(x/y)$. This is true for all probabilities with low mutual information with the halting sequence. The function $\K$ is the prefix-free Kolmogorov complexity.

To prove this fact, we first prove an interesting property about bunches of finite strings. A $(k,l)$-bunch is a finite set of strings $X$ where $l>\max_{x,y\in X} \K(y/x)$ and $2^k<|X|$. Bunches were introduced in \citep{Romashchenko03}, but we use a slightly different definition. Although bunches have only two parameters, they exhibit many interesting properties. Both \citep{Romashchenko03} and \citep{Romashckenko22} proved the existence of strings that are simple to each member of the bunches. That is, there exists a string $z$ such that $\K(z/x)<O(l-k)+\K(l)$ and $\K(x/z)<l+O(l-k)+\K(l)$, for all $x\in X$. In \citep{EpsteinBunch21}, it was proven that each bunch has a member that is simple relative to all members of the bunch, similar to the above definition.
If not, then the bunch has high mutual information with the halting sequence. The mutual information between a string and the halting sequence is $\I(x;\mathcal{H})=\K(x)-\K(x/\mathcal{H})$. 
We prove that if a nonexotic bunch $X$ has many members and low $\max_{x,y\in X,x\neq y}\K(y/x)$, then it will have two elements $x,y$ with very low $\K(y/x)$. A string (or any object that it is represented by) is exotic if it has high mutual information with the halting sequence. \\

\noindent\textbf{Theorem. }\textit{For $(k,l)$-bunch $X$, $\min_{x,y\in X, x\neq y}\K(y/x)\lel \ceil{l-2k}^++\I(X;\ch)+2\K(k,l)$.}\\

\noindent\textbf{The Kolmogorov Birthday Paradox}. Let us say we select a random subset $D$ of size $2^{n/2}$ consisting of (possibly repeated) strings of length $n$, where each string is selected independently with a uniform probability. For the simple Kolmogorov birthday paradox, with overwhelming probability, there are two (possibly the same) strings $x,y\in D$, such that $\K(x/y)=O(1)$, for a large enough constant. This is due to reasoning from the classical birthday paradox. We now prove the general Kolmogorov birthday paradox. Let $P$ be any probability over sets $D$ consisting of $2^{n/2}$ (non repeated) strings of length $n$. Since $D\subset \BT^n$, for all $D$, $\max_{x,y\in D}\K(x/y)\lea n$. By Corollary \ref{cor:consinf} in Appendix \ref{sec:cons}, $\Pr_{D\sim P}\left[\I(D;\ch)>\I(P;\ch)+m\right]\lem 2^{-m}$. Combining these facts with the above theorem, with $l=n+O(1)$ and $k=.5n-1$, we obtain the following result.  \\

\noindent\textbf{Corollary. } $\Pr_{D\sim P}\big[\min_{x,y\in D, x\neq y}\K(x/y)\lel \I(P;\ch)+2\K(n)+c\big] > 1-2^{-c}$.\\

Obviously, the bound loosens if $P$ samples sets of smaller size, mirroring the classical birthday paradox.
\section{Related Work}

The study of Kolmogorov complexity originated from the work of~\citep{Kolmogorov65}. The canonical self-delimiting form of Kolmogorov complexity was introduced in~\citep{ZvonkinLe70} and treated later in~\citep{Chaitin75}. The universal probability $\m$ was introduced in~\citep{Solomonoff64}. More information about the history of the concepts used in this paper can be found in textbook~\citep{LiVi08}. 

The main result of this paper is an inequality including the mutual information of the encoding of a finite set with the halting sequence. A history of the origin of the mutual information of a string with the halting sequence can be found in~\citep{VereshchaginVi04v2}.

A string is stochastic if it is typical of a simple elementary probability distribution. A string is typical of a probability measure if it has a low deficiency of randomness. The deficiency of randomness of a number $a\in\N$ with respect to a probability $P$ is $\d(a|P)=-\log P(a)-\K(a/\langle P\rangle)$. It is a measure of the extent of the refutation against the hypothesis $P$ given the result $a$ \cite{Gacs21}. Thus, the stochasticity, $\Ks(a)$, of a string $a$ is roughly $\min_{\textrm{ probability }P}\K(P) + O(\log \d(a|P))$.

In the proof of Theorem \ref{thr:graph}, the stochasticity measure of encodings of finite sets is used. The notion of the deficiency of randomness with respect to a measure follows from the work of~\citep{Shen83} and is also studied in~\citep{KolmogorovUs87,Vyugin87,Shen99}. Aspects involving stochastic objects were studied in~\citep{Shen83,Shen99,Vyugin87,Vyugin99}. 

	\begin{figure}               
  \centering
  \includegraphics[width=13cm]{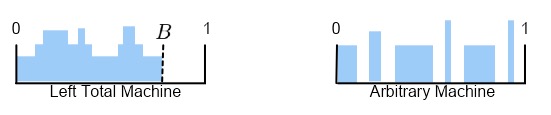}
\caption{The domain of a Turing machine $T$ can be interpreted as the $[0,1]$ interval, and the strings for which $T$ halts can be seen as a collection of dyadic subintervals. A left-total machine $L$ has the property that if $L$ halts on a string $x$, then it will halt on a string $y$ whose binary interval is smaller (i.e., to the left of $x$). The infinite sequence $B$ is called the border sequence and is the binary expansion of Chaitin's Omega. This paper uses a left-total universal Turing machine.}
  \label{fig:lefttotal}
\end{figure}

This work uses the notion of left-total machine (see Figure \ref{fig:lefttotal}) and the notion of the infinite ``border'' sequence, which is equal to the binary expansion of Chaitin's Omega (see Section \ref{sec:LeftTotal}). The works of ~\citep{VereshchaginVi04v2,GacsTrVi01} introduced the notion of using the prefix of the border sequence to define strings into a two-part code.
This paper uses the lemmas found in \citep{Epstein21}.

This paper can be seen as a conditional variant to the main result in \cite{Levin16}. \cite{Levin16} proved that for nonexotic sets $D$, the a priori probability, $\m$, of a set is concentrated on a single element.\\

\noindent \textbf{Theorem.} (\cite{Levin16}) $-\log \max_{x \in D}\m(x)\lel-\log \sum_{x \in D}\m(x)+ \I(D;\ch)$.\\

There is a simple proof for this theorem in \cite{Shen12}. The proof of Theorem \ref{thr:graph} is similar to that of the main result in \cite{Levin16}, in that they both first prove stochasticity, $\Ks(O)$, of an object $O$ with certain properties and then show that this object has high $\I(O;\ch)$. In \cite{Levin16}, $O$ is equal to a set, and in this paper, $O$ is equal to a (sub)graph. Theorem \ref{thr:klset} is not directly implied by the theorem in \cite{Levin16} because this paper addresses conditional complexities between elements of a set. In addition, Theorem \ref{thr:klset} is not a generalization of the main theorem in \cite{Levin16} because it relies on the parameters of bunches and not the a priori probability $\m$.

\section{Conventions}

\label{sec:conv}
We use $\BT$, $\FS$, $\IS$ $\mathbb{W}$, $\N$, $\Q$, and $\R$ to denote bits, finite strings, infinite sequences, whole numbers, natural numbers, rationals, and reals, respectively. Let $X_{\geq 0}$ and $X_{> 0}$ be the sets of nonnegative and positive elements of $X$. $\FIS=\FS\cup\IS$. The positive part of a real is $\ceil{a}^+=\max\{a,0\}$. For string $x\in\FS$, $x0^-=x1^-=x$. For $x\in\FS$ and $y\in\FIS$, we use $x\sqsubseteq y$ if there is some string $z\in\FIS$ where $xz=y$. We say $x\sqsubset y$ if $x\sqsubseteq y$ and $x\neq y$. The indicator function of a mathematical statement $A$ is denoted by $[A]$, where if $A$ is true, then $[A]=1$; otherwise, $[A]=0$. The self-delimiting code of a string $x\in\FS$ is $\langle x\rangle = 1^{\|x\|}0x$. The encoding of (a possibly ordered) set $\{x_1,\dots,x_m\}\subset\FS$ is $\langle m\rangle\langle x_1\rangle\dots\langle x_m\rangle$. 

Probability measures $Q$ over numbers are elementary if $|\mathrm{Support}(Q)|<\infty$ and $\mathrm{Range}(Q)\subset Q_{\geq 0}$. Elementary probability measures $Q$ with $\{x_1,\dots,x_m\}=\mathrm{Support}(Q)$ are encoded by finite strings, with $\langle Q\rangle = \langle\{x_1,Q(x_1),\dots,x_m,Q(x_m)\}\rangle$. For the nonnegative real function $f$, we use $\lea f$, $\gea f$, and $\eqa f$ to denote $<f+O(1)$, $>f-O(1)$, and $=f\pm O(1)$. We also use $\lel f$ and $\gel f$ to denote $<f + O(\log (f+1))$ and $>f - O(\log (f+1))$, respectively.

We use a universal prefix-free algorithm $U$, where we say $U_\alpha(x)=y$ if $U$, on main input $x$ and auxiliary input $\alpha$, outputs $y$. We define Kolmogorov complexity with respect to $U$, where if $x\in\FS$, $y\in\FIS$, then $\K(x/y)=\min\{\|p\|:U_y(p)=x\}$. The universal probability $\m$ is defined as $\m(x/y)=\sum_p[U_y(p)=x]2^{-\|p\|}$. By the coding theorem, $\K(x/y)\eqa-\log\m(x/y)$. By the chain rule, $\K(x,y)\eqa\K(x)+\K(y/x,\K(x))$. The halting sequence $\ch\in\IS$ is the unique infinite sequence where $\ch[i]=[U(i)\textrm{ halts}]$. The information that $x\in\FS$ has about $\ch$, conditional on $y\in\FIS$, is $\I(x;\ch/y)=\K(x/y)-\K(x/\langle y,\ch\rangle)$. $\I(x;\ch)=\I(x;\ch/\emptyset)$.
 
This paper uses notions of stochasticity in the field of algorithmic statistics \citep{VereshchaginSh17}. A string $x$ is stochastic, i.e., has a low $\Ks(x)$ score if it is typical of a simple probability distribution. The extended deficiency of the randomness function of a string $x$ with respect to an elementary probability measure $P$ conditional on $y\in\FS$ is $\d(x|P,y)=\floor{-\log P(x)}-\K(x/\langle P\rangle,y)$. $\d(x|P) = \floor{-\log P(x)}-\K(x/\langle P\rangle)$ 
 \begin{dff}[Stochasticity]
 	For $x,y\in\FS$, $\Ks(x/y)=\min \{\K(P/y)+3\log \max\{\d(x|P,y),1\}:$ $ P\textrm { is an elementary probability measure}\}$. $\Ks(x) = \Ks(x/\emptyset)$.
 \end{dff}
\section{Labeled Graph, Warm Up}
In Section \ref{sec:graph}, a property of a complete subgraph of a labeled graph is proven. A labeled graph is a directed graph such that each vertex has a unique string attached to it. Given certain properties of the graph $G=(G_E,G_V)$, where $G_E$ are the directed edges, $G_V$ are the vertices, and subgraph $J=(J_E,V_V)$, Theorem \ref{thr:graph} in Section \ref{sec:graph} proves that $J$ is guaranteed to have an edge $(x,y)\in J_E$ with low $\K(x|y)$. In this section, we describe the overall arguments in the proof of this theorem.

We specify a vertex interchangeably with the string assigned to it. The general argument for the proof of Theorem \ref{thr:graph} is as follows. Given a labeled graph $G$, if there is a random subgraph $F=(F_E,F_V)$ that is large enough, then it will probably share an edge with most large complete subgraphs $J$ of $G$. Thus, large complete subgraphs of $G$ with an empty intersection with $F$ will be considered atypical. If $F$ shares an edge with complete subgraph $J\subseteq G$, then 
$$\min_{(x,y)\in J_E}\K(y/x)\lessapprox\log \max_{x\in F_V}\mathrm{OutDegree}(x)+\K(F).$$

This inequality follows from the fact that given a description of $F$ describing $\{(x,y):(x,y)\in F_E\}$ and an $x\in F$, each $y\in \{y:(x,y)\in F_E\}$ can be described relative to $x$ with $\ceil{\log\ \mathrm{OutDegree}(x)}$ bits. In this section, instead of using random subgraphs, we use random lists of vertices $L_\bullet$, indexed by $x\in G_V$. Thus, for each $x\in G_V$, $L_x$ is a list of vertices, possibly with repetition. This allows for easier manipulation.

The warm-up arguments are as follows. Let $G=(G_E,G_V)$ be a graph of max degree $2^l$ and $\mathcal{J}$ be the set of complete subgraphs of $G$ of size $2^k$. We assume $l>2k$. Each vertex $x\in G_V$ has a random list $L_x$ of $2^{l-2k}$ vertices, where for $i\in[1,2^{l-2k}]$, $\Pr(y= L_x[i])=[(x,y)\in G_E]2^{-l}$ and $\Pr(\emptyset=L_x[i])=1-\mathrm{OutDegree}(x)2^{-l}$. For $J\in\mathcal{J}$, indexed list $L_\bullet$, 
$$\mathrm{Miss}(J,L_\bullet)\textrm{ is true iff }\forall x,\forall y\in J_V, y\not\in L_x.$$ 
For each $J\in\mathcal{J}$, 
\begin{align*}
\Pr(\mathrm{Miss}(J,L_\bullet))& = \prod_{x\in J_V}\Pr(\forall y\in J_V,y\not\in L_x) \\
&\leq \prod_{x \in J_V} \prod_{i \in [1, 2^{l-2k}]} \Pr [\forall y \in J_V, y \neq L_x[i]]\\
&\leq \prod_{x\in J_V}\left(1-2^{k-l}\right)^{|L_x|}\\
&\leq \prod_{x\in J_V}\left(1-2^{k-l}\right)^{2^{l-2k}}\\
&\leq \left((1-2^{k-l})^{2^{l-2k}}\right)^{|J|}\\
& \leq {\left(e^{-2^{-k}}\right)}^{|J|} < e^{-1}< 1.
\end{align*}
Now assume that $|L_x| = b2^{l-2k}$ for all $x \in G_V$, i.e., $b$ times more than before. It is not hard to see that $\Pr(\mathrm{Miss}(J,L_\bullet))< e^{-b}$ for each $J\in\mathcal{J}$. We assume a uniform distribution $\mathcal{U}$ over $\mathcal{J}$ (i.e. complete subgraphs of size $2^k$). Under this assumption,
\begin{align*}
\E\Big[[\mathrm{Miss}(J,L_\bullet)]\Big]<\sum_{J\in\mathcal{J}} |\mathcal{J}|^{-1}e^{-b}= e^{-b}.
\end{align*}
Thus, given all the parameters, $G$, $k$, $l$, and $b$, using brute force search, one can find a set of lists $L'_\bullet$ of size $b2^{l-2k}$ indexed by $x\in G_V$, such that less than $e^{-b}$ of members $J$ of $\mathcal{J}$ have $\mathrm{Miss}(J,L'_\bullet)$. If $\mathrm{Miss}(J,L_\bullet')$ is true for $J\in \mathcal{J}$, then it must be atypical of $\mathcal{U}$ because $\E_{J\sim\mathcal{U}}[\mathrm{Miss}(J,L_\bullet')]<e^{-b}$. One can construct a $\mathcal{U}$-test using $L'_\bullet$. A $\mathcal{U}$-test is any function $t:\FS\rightarrow \R_{\geq 0}$ such that $\sum_{J\in\mathcal{J}}t(J)\mathcal{U}(J)\leq 1$. Thus, $t\cdot \mathcal{U}$ is a semimeasure, and therefore,
\begin{align}
\label{eq:testsemi}
\K(J/t,\mathcal{U}) \lea -\log t(J)\mathcal{U}(J).
\end{align}
Thus, the function $t(J)=[\mathrm{Miss}(J,L_\bullet')]e^b$ is a $\mathcal{U}$-test, with $\sum_{J\in\mathcal{J}}t(J)\mathcal{U}(J)<1$. We set aside the parameters $(G,k,l,b,\mathcal{U})$ because they complicate the discussion. That is, we roll the parameters into the additive constants of the inequalities.
By the definition of randomness deficiency,

\begin{align}
\nonumber
\d(J|\mathcal{U})&=-\log \mathcal{U}(J)-\K(J/\mathcal{U})\\
\label{eq:DefJ}
&\gea \log |\mathcal{J}| - \K(J/L'_\bullet)\\
\label{eq:addT}
&\gea \log |\mathcal{J}| - \K(J/t)\\
\label{eq:applytestsemi}
&\gea \log |\mathcal{J}| + \log t(J)\mathcal{U}(J)\\
\nonumber 
&\gea \log |\mathcal{J}| + \log t(J)|\mathcal{J}|^{-1}\\
\nonumber
&\gea b\log e.
\end{align}
Equation \ref{eq:DefJ} has two components. The first term $\log |\mathcal{J}|$ is equal to $-\log\mathcal{U}(J)$ because $\mathcal{U}$ is the uniform distribution over all $\mathcal{J}\ni J$, the set of all complete subgraphs of $G$ of size $2^k$. The second term is due to the additive equalities $$\K(J/L'_\bullet)=\K(J/L'_\bullet,G,k,l,b,\mathcal{U})\eqa \K(J/G,k,l,b,\mathcal{U})\eqa\K(J/\mathcal{U}),$$ 
in that given all the hidden parameters $(G,k,l,b,\mathcal{U})$, one can compute $L'_\bullet$ using brute force search, as described above. Equation \ref{eq:addT} derives from the test $t$ being constructed from $L'_\bullet$ (and the hidden parameters). Equation \ref{eq:applytestsemi} is due to the properties of the tests, as shown in Equation \ref{eq:testsemi}.

Thus, all complete subgraphs $J\in\mathcal{J}$ of $G$ for which $\mathrm{Miss}(J,L'_\bullet)$ is true will be atypical of $\mathcal{U}$, with randomness deficiency $\d(J|\mathcal{U})$ greater than $b$. Thus, if a subgraph $J\in\mathcal{J}$ is $b$-typical, then there exists $(x,y)\in J_E$, with $y\in L_x$. Therefore, $b$-typical subgraphs $J\in\mathcal{J}$ will have
\begin{align}
\label{eq:warmup1}
\min_{(x,y)\in J_E}\K(y/x)&\lea\log \big|L_x\big|\lea  l-2k+\log b.
\end{align} 
For Theorem \ref{thr:graph}, the uniform probability measure $\mathcal{U}$ is replaced by a special computable measure $P$ that realizes the stochasticity $\Ks$ of the subgraph $J$. In addition, $b$ is chosen to equal $b\approx \d(J|P)$ so that the subgraph $J$ is guaranteed to be typical of $P$, so $\mathrm{Miss}(J)$ is false. This means that Equation \ref{eq:warmup1} holds for $J$. In addition, in the next section, the parameters $(G,k,l,b)$ must be taken into account. 
\section{Labeled Graphs}
\label{sec:graph}
In this section, we study exotic subgraphs of simple labeled graphs. A subgraph $J$ is exotic if consists of labeled edges $(x,y)\in J_E$, such that the conditional complexity $\K(y/x)$ is high. The proof of the following theorem uses stochasticity $\Ks$.  
An example proof that uses $\Ks$ and mirrors the proof of Theorem \ref{thr:graph} can be found in Appendix \ref{sec:exerstoc}. Note that the lemma in Appendix \ref{sec:exerstoc} is just an exercise to demonstrate reasoning with $\Ks$. The lemma is not used in the paper.
\begin{thr}
	\label{thr:graph}
	\noindent For graph $G=(G_E,G_V)$, complete subgraph $J=(J_E,J_V)$; if $2^l> \max\mathrm{Outdegree}(G)$, $2^k< |J|$, then we have $\min_{(x,y)\in J_E}\K(y/x)\lel \ceil{l-2k}^++\I(J;\ch/G,k)+\K(G,k)$.
\end{thr}
\begin{prf}
We put $( G,k)$ on an auxiliary tape to the universal Turing machine $U$. Thus, all algorithms have access to $(G,k)$, and all complexities implicitly have $(G,k)$ as conditional terms.

Let $\ell=\max\{l,2k\}$. Let $P$ be the probability that realizes $\Ks(J)$ and the deficiency of randomness $d=\max\{\d(J|P),1\}$. Let $V:G\times G\rightarrow\mathbb{R}_{\geq 0}$ be a conditional probability measure where $V(y|x)=[(x,y)\in G_E]2^{-\ell}$ and $V(\emptyset|x)=1-\mathrm{OutDegree}(x)2^{-\ell}$. We define a conditional probability measure over lists $L$ of $cd2^{\ell-2k}$ vertices of $G$, with $\kappa : G\times G^{cd2^{\ell-2k}}\rightarrow\mathbb{R}_{\geq 0}$, where $\kappa(L|x)=\prod_{y\in L}V(y|x)$. The constant $c\in\N$ will be determined later. Let $L_{\bullet}$ be an indexed list of $cd2^{\ell-2k}$ elements, indexed by $x\in G$, where each list is denoted by $L_x$ for $x\in G_V$. Let $\kappa(L_\bullet)=\prod_{x\in G}\kappa(L_x|x)$. A graphical representation of $\kappa$ and $L_\bullet$ can be found in Figure \ref{fig:L}.\begin{figure}     
  \centering
  \includegraphics[width=6cm]{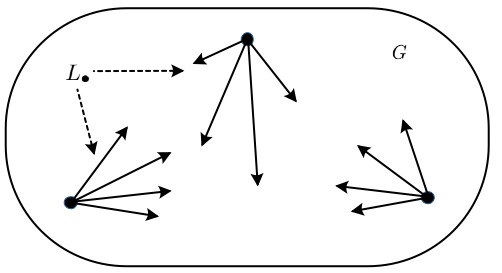}
  \caption{The above diagram is a graphical representation of $\kappa$ and $L_\bullet$, assuming that $cd2^{\ell-2k}=4$. Each vertex has four edges chosen at random, where each particular edge is chosen with probability $2^{-\ell}$.}
  \label{fig:L}
\end{figure} For indexed list $L_\bullet$ and graph $H=(H_E,H_V)$, we use the indicator $\mathbf{i}(L_\bullet,H)=[\textrm{Complete }H\subseteq G, 2^k< |H_V|, \forall (x,y)\in H_E, y\not\in L_x]$.
	\begin{align*}
	\E_{L_\bullet\sim \kappa}\E_{H\sim P}[\i(L_\bullet,H)]&\leq \sum_HP(H)\Pr_{L_\bullet\sim \kappa}(\forall (x,y)\in H_E, y\not\in L_x, |H_V|>2^k,\textrm{ Complete }H\subseteq G)\\
	&\leq \sum_HP(H)[|H_V|>2^k]\prod_{x\in H_V}(1-2^{k-\ell})^{|L_x|}\\
	&\leq \sum_HP(H)[|H_V|>2^k]\prod_{x\in H_V}(1-2^{k-\ell})^{cd2^{\ell-2k}}\\
	&\leq \sum_HP(H)[|H_V|>2^k]\prod_{x\in H_V}e^{-cd2^{-k}}\\
	&< \sum_HP(H)e^{-cd}\\
	&= e^{-cd}. 
	\end{align*}
	Thus, there exists an $L'_\bullet$ such that $\E_{H\sim P}[\i(L'_\bullet,H)]< e^{-cd}$. This $L'_\bullet$ can be found with brute force search with all the parameters, with 
\begin{align}
\label{eq:LSimple}
\K(L'_\bullet/P,c,d)&=O(1). 
\end{align}
Thus, $t(H)=\i(L'_\bullet,H)e^{cd}$ is a $P$ test, where $\E_{H\sim P}[t(H)]\leq 1$. This test $t$ gives a high score to complete subgraphs of $G$ of size $>2^k$ that have no intersecting edges with $L'_\bullet$. A diagram of the components used in this proof can be found in Figure \ref{fig:graph}. Furthermore, 
	\begin{align*}
	\K(t|P,c,d)&\eqa\K(t|L'_\bullet, P,c,d)=O(1).
	\end{align*}
\begin{figure}[t!]      
  \centering
  \includegraphics[width=13cm]{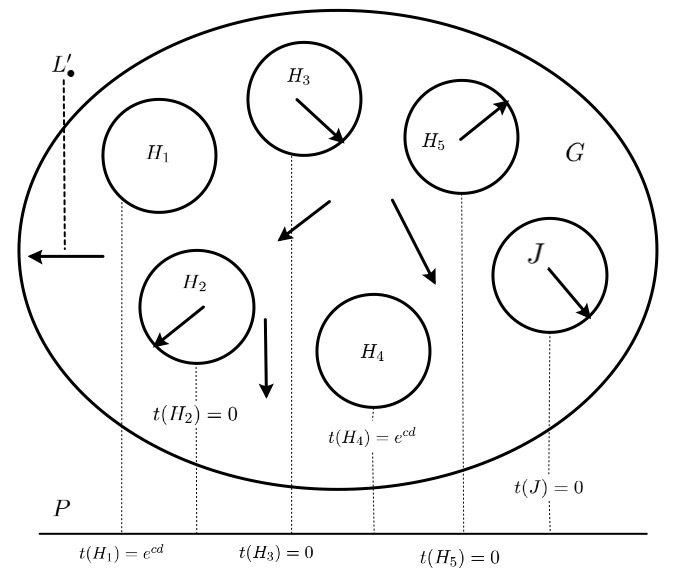}
\caption{The above diagram is a graphical representation of the concepts used in the proof of Theorem \ref{thr:graph}. The main ellipse models the graph $G$, and the circles in the graph represent complete subgraphs (labeled $H_1$ to $H_5$ and $J$) with $>2^k$ vertices. Each subgraph is in the support of probability $P$, represented by the dotted lines. The set $L'_\bullet$ represents a collection of selected edges. If a subgraph $H_i$ does not contain an edge in $L'_\bullet$, then $H_i$ is \textit{atypical} and has a high score $t(H_i)$. By design, $J$ is typical and thus shares an edge with $L'_\bullet$.}
  \label{fig:graph}
\end{figure}It must be that there is an $(x,y)\in J_E$ where $y\in L_x$. Otherwise, $t_{L_\bullet}(J)=e^{cd}$ and 
	\begin{align}
	\nonumber
	\K(J/P,c,d) &\lea \K(J/t,P,c,d)\\
	\label{eq:ApplyTestGraph}
	\K(J/P,c,d) &\lea -\log t(J)P(J)\\
	\nonumber
	&\lea -(\log e) cd-\log P(J)\\
	\nonumber
	(\log e) cd&\lea -\log P(J)  - \K(J/P,c,d)\\
	\nonumber
	(\log e) cd&\lea -\log P(J)  - \K(J/P)+\K(c,d)\\
	\nonumber
	(\log e) cd&\lea d+\K(c,d),
	\end{align}
	which is a contradiction for large enough $c$ solely dependent on the universal Turing machine $U$. Equation \ref{eq:ApplyTestGraph} is due to Equation \ref{eq:testsemi}. The constant $c$ is folded into the additive constants of the inequalities of the rest of the proof. Thus, since there exists $(x,y)\in J_E$ where $y\in L_x$,
	\begin{align}
	\nonumber
	\K(y/x) &\lea \log |L'_x|+\K(L'_\bullet)\\
	\nonumber
	&\lea\ceil{l-2k}^++\log d + \K(L'_\bullet/P,d)+\K(P,d)\\
	\label{eq:applylsimple}
	&\lea\ceil{l-2k}^++\log d +\K(P,d)\\
	\nonumber
	&\lea\ceil{l-2k}^++3\log d +\K(P)\\
	\label{eq:defks}
	&\lea\ceil{l-2k}^++\Ks(D)
	\end{align}
		Equation \ref{eq:applylsimple} is due to Equation \ref{eq:LSimple}. Equation \ref{eq:defks} is due to the definition of stochasticity. We now make the relativization of $(G,k)$ explicit, with
	\begin{align}
	\nonumber
	\K(y/x,G,k) &\lea\ceil{l-2k}^++\Ks(J/G,k)\\
	\label{eq:stochh}
	&\lel\ceil{l-2k}^++\I(J;\mathcal{H}/G,k)\\
	\nonumber
	\K(y/x) & \lel\ceil{l-2k}^++\I(J;\mathcal{H}/G,k)+\K(G,k).
	\end{align} Equation \ref{eq:stochh} is due to Lemma 10 in \cite{Epstein21}, which states $\Ks(x)<\I(x\,{:}\,\ch)+O(\K(\I(x\,{:}\,\ch)))$.\qed\\
\end{prf}

\section{Warm Up for the Main Theorem of the Paper}
Theorem \ref{thr:graph} can be used to prove results about the minimum conditional complexity between two elements of a bunch. This section gives a broad overview of the arguments used in the proof of Theorem \ref{thr:klset}. Let $X\subset\FS$ be a $(k,l)$-bunch, where $|X|> 2^k$, and $\max_{x,y\in X}\K(y/x)< l$. 

Let $\K^r(x/y)=\min\{\|p\|:U_y(p)=x\textrm{ in time }r\}$ be the conditional complexity of $x$ given $y$ at time $r$. Therefore, given a number $r$, $\K^r$ is computable. We also assume $\K^r(x/y)=\infty$ if $\|y\|>r$ to ensure that $\K^r$ has finite $\{(x,y) : \K^r(x/y)<\infty, x,y\in\N\}$ for each $r$. Let $G^r=(G^r_E,G^r_V)$ be a graph defined by $(x,y)\in G^r_E$ iff $\K^r(x/y)<l$.

Let $s$ be the smallest number where $\K^s(x/y)< l$, for all $x,y\in X$. Let $G=(G_E,G_V)=G^s$.  Since $X$ is a $(k,l)$-bunch, $X$ can be viewed as a complete subgraph of $G$ of size ${>}2^k$. Invoking Theorem \ref{thr:graph}, we obtain
\begin{align}
\label{eq:approxinequality}
\min_{(x,y)\in X, x\neq y}\K(y/x)&\lel \ceil{l-2k}^++\I(X;\mathcal{H}/G,k)+\K(G,k).
\end{align}
We have $\K(s/G)\lea \K(l)$ because $s=\min\{r:G=G^r\}$. Therefore,
\begin{align}
\label{eq:XGk}
\K(X/G)&\lea \K(X/s)+\K(s/G)\lea \K(X/s)+\K(l).
\end{align} 
Due to the definition of $G=G^s$,
\begin{align}
\label{eq:Gs}
\K(G/s)&\lea \K(l).
\end{align}
 By the definition of $\I$,
\begin{align}
\nonumber
\I(X;\mathcal{H}/G,k)&=\K(X/G,k)-\K(X/G,k,\mathcal{H})\\
\nonumber
&=\K(X/G)-\K(X/G,\mathcal{H})+O(\K(k))\\
\label{eq:applyXKg}
&\lea \K(X/s)-\K(X/G,\mathcal{H})+O(\K(k,l))\\
\nonumber
&< \K(X/s)-\K(X/s,\mathcal{H})+\K(G/s)+O(\K(k,l))\\
\label{eq:xgr}
&< \I(X;\mathcal{H}/s)+O(\K(k,l)).
\end{align}
Equation \ref{eq:applyXKg} is due to Equation \ref{eq:XGk}. Equation \ref{eq:xgr} is due to Equation \ref{eq:Gs}. Using $\K(G)\lea\K(s)+\K(l)$ and Equation \ref{eq:xgr}, we obtain
\begin{align}
\label{eq:warmpen}
\I(X;\mathcal{H}/G,k)+\K(G,k) &< \I(X;\ch/s) +\K(s) + O(\K(k,l)).
\end{align}
Combining Equations \ref{eq:approxinequality} and \ref{eq:warmpen},
we obtain 
\begin{align}
\label{eq:approxinequalityr}
\min_{(x,y)\in J_E}\K(y/x)&\lel \ceil{l-2k}^++\I(X;\mathcal{H}/s)+\K(s)+O(\K(k,l)).
\end{align}
This inequality is close to the form of Theorem \ref{thr:klset}. The main difference is that the number $s$ appears in Equation \ref{eq:approxinequalityr}. This can be rectified if we use a different notion of a computational resource. In the next section, we introduce left-total universal machines, and the resource used is not a number $s$ but a so-called total string $b$. Then, Lemma \ref{lmm:totalString}, defined in Section \ref{sec:LeftTotal}, can be used to remove the $b$ factor from the final inequality.

\section{Left-Total Machines}
\label{sec:LeftTotal}
\begin{figure}[h!]
	\begin{center}
		\includegraphics[width=0.5\columnwidth]{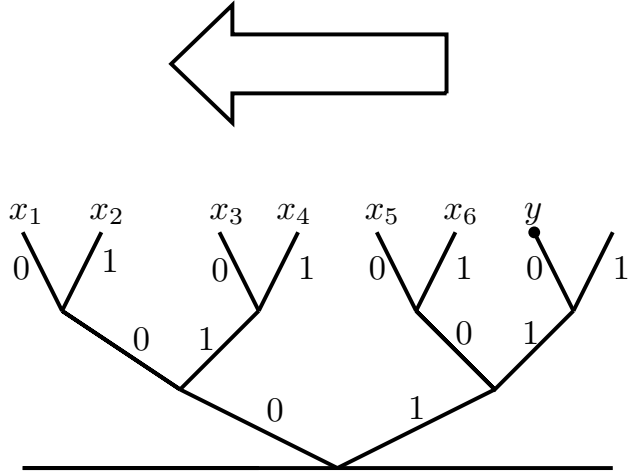}
		\caption{The above diagram represents the domain of a left-total machine $T$ with the 0 bits branching to the left and the 1 bits branching to the right. For $i\in \{1,\dots,5\}$, $x_i\lhd x_{i+1}$ and $x_i\lhd y$. Assuming $T(y)$ halts, each $x_i$ is total. This also implies that each $x_i^-$ is total.}
		\label{fig:LeftTotal}
	\end{center}
\end{figure}

We recall that for $x\in\FS$, $\Gamma_x=\{x\beta:\beta\in\IS\}$ is the interval of $x$.
The notions of total strings and the ``left-total'' universal algorithm are needed in this paper. We say $x\in\FS$ is total with respect to a machine if the machine halts on all sufficiently long extensions of $x$. More formally, $x$ is total with respect to $T_y$ for some $y\in\FIS$ iff there exists a finite prefix-free set of strings $Z\subset\FS$ where $\sum_{z\in Z}2^{-\|z\|}=1$ and $T_y(xz)\neq\perp$ for all $z\in Z$. We say (finite or infinite) string $\alpha\in\FIS$ is to the ``left'' of $\beta\in\FIS$ and use the notation $\alpha\lhd \beta$ if there exists an $x\in\FS$ such that $x0\,{\sqsubseteq}\, \alpha$ and $x1\,{\sqsubseteq}\, \beta$. A machine $T$ is left-total if for all auxiliary strings $\alpha\in\FIS$ and for all $x,y\in\FS$ with $x\lhd y$, one has that $T_\alpha(y)\neq\perp$ implies that $x$ is total with respect to $T_\alpha$. Left-total machines were introduced in \citep{Levin16}. An example can be seen in Figure \ref{fig:LeftTotal}.

For the remainder of this paper, we can and will change the universal self-delimiting machine $U$ into a universal left-total machine $U'$ by the following definition. The algorithm $U'$ orders all strings $p\,{\in}\,\FS$ by the running time of $U$ when given $p$ as an input. Then, $U'$ assigns each $p$ an interval $i_p{\subseteq}[0,1]$ of width $2^{-\|p\|}$. The intervals are assigned ``left to right'', where if $p\in\FS$ and $q\in\FS$ are the first and second strings in the ordering, then they will be assigned the intervals $[0,2^{-\|p\|}]$ and $[2^{-\|p\|},2^{-\|p\|}\,{+}\,2^{-\|q\|}]$, respectively. 

Let the target value of $p\in\FS$ be $(p)\,{\in}\,\mathbb{W}$, which is the value of the string in binary. For example, the target value of both strings 011 and 0011 is 3. The target value of 0100 is 4. The target interval of $p\in\FS$ is $\Gamma(p)=((p)2^{-\|p\|},((p){+}1)2^{-\|p\|})$.

The universal machine $U'$ outputs $U(p)$ on input $p'$ if the intervals $\Gamma(p')$ are strictly contained in $i_p$, with $\Gamma(p')\subset i_p$, and $\Gamma({p'}^{-})$ are not strictly contained in $i_p$, with $\Gamma({p'}^{-})\not\subset i_p$. The same definition applies to machines $U'_\alpha$ and $U_\alpha$ over all $\alpha\,{\in}\,\FIS$. 

Recall that a function $f:\N\rightarrow\N$ is partially computable with respect to $U$ if there is a string $t\in\FS$ such that $f(x)=U(t\langle x\rangle)$ when $f(x)$ is defined and $U(t\langle x\rangle)$ does not halt otherwise. Similarly, a function $f:\N\rightarrow\N$ is partially computable with respect to $U'$ if there is $t\in\FS$, such that whenever $f(x)$ is defined, there is an interval $i_{t\langle x\rangle}$ and for any string $p$ where $\Gamma(p)$ and not that of $\Gamma(p^-)$ is contained in $i_{t\langle x\rangle}$, then $U'(p)=f(x)$. Otherwise, if $f(x)$ is not defined, the interval $i_{t\langle x\rangle}$does not exist. The following proposition was used without being proven in \citep{Levin16}.

\begin{prp}
	$\K_U(x/y)\eqa\K_{U'}(x/y)$.
	\label{prp:kequiv}
\end{prp}
\begin{prf}
	It must be that $\K_U(x/y)\lea \K_{U'}(x/y)$ because there is a Turing machine that computes $U'$. Therefore, due to the universality of $U$, there is a $t\in\FS$, such that $U_y(tx)=U'_y(x)$, thus proving the minimality of $\K_U$. It must be that $\K_{U'}(x/y)\lea \K_U(x/y)$. This is because if $U(x)=z$, then there is interval $i_x$ such that for all strings $p$ where $\Gamma(p)$ and not that of $\Gamma(p^-)$ that are strictly contained in $i_x$ has $U'_y(p)= U_y(x)$. Thus, we have that $\|p\|\leq \|x\|+2$. This implies that $\K_{U'}(x/y)\leq \K_U(x/y)+2$.\qed
\end{prf}

For the rest of the paper, we now set $U$ to be equal to $U'$, so the universal Turing machine can be considered to be left-total. Without loss of generality, as shown in Proposition \ref{prp:kequiv}, the complexity terms of this paper are defined with respect to the universal left-total machine $U$.

\begin{prp}
\label{prp:unique}
	There exists a unique infinite sequence $\mathcal{B}$ with the following properties. 
	\begin{enumerate}
		\item All the finite prefixes of $\mathcal{B}$ have total and nontotal extensions. 
		\item If a finite string has total and nontotal extensions, then it is a prefix of $\mathcal{B}$.
		\item  If a string $b$ is total and $b^-$ is not, then $b^-\sqsubset\mathcal{B}$.
	\end{enumerate}
\end{prp}
\begin{prf}
\begin{enumerate}
\item Let $\Omega\in\R$ be Chaitin's Omega, the probability that a random sequence of bits halts when given to $U$, with $\Omega =\sum_{p\in\FS}[U(p)\textrm{ halts}]2^{-\|p\|}$. Thus, $\Omega$ characterizes the domain of $U$, with $\bigcup_{p\in\FS}i_p=[0,\Omega)$. Let $\mathcal{B}\in \IS$ be the binary expansion of $\Omega$, which is an 
ML
 random string. For each $n\in \N$, let $b_n\sqsubset\mathcal{B}$, $\|b_n\|=n$. Let $m\in\mathbb{W}$ be the smallest whole number such that $b_n1^{(m)}0\sqsubset\mathcal{B}$. Then, $b_n1^{(m+1)}$ is a nontotal string because $[0,\Omega]\cap\Gamma(b_n1^{(m+1)})=\emptyset$. Furthermore, let $m\in\mathbb{W}$ be the smallest whole number such that $b_n0^{(m)}1\sqsubset\mathcal{B}$. Then, $b_n0^{(m+1)}$ is a total string because $\Gamma(b_n0^{(m+1)})\subset[0,\Omega)$.
\item Assume there are two strings $x$ and $y$ of length $n$ that have total and nontotal extensions, with $x\lhd y$. Since $y$ has total extensions, there exist $z$ such that $U'(yz)$ halts. Since $x\lhd yz$, by the definition of left-total machines, $x$ is total, causing a contradiction.
\item This is because $b^-$ has total and nontotal extensions.
\end{enumerate}
 \qed\\
\end{prf}

\begin{figure}[h!]
	\begin{center}
		\includegraphics[width=0.4\columnwidth]{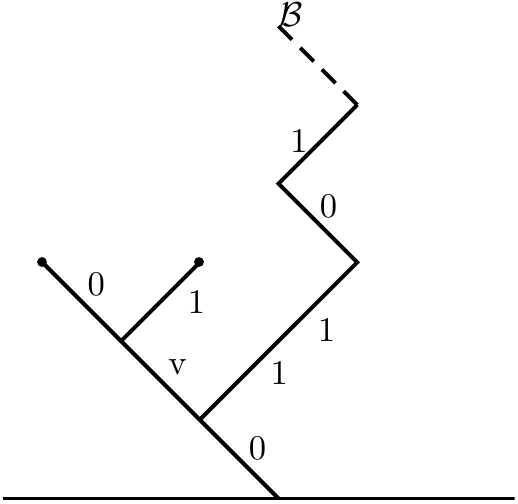}
		\caption{The above diagram represents the domain of the universal left-total algorithm $U'$, with 0 bits branching to the left and 1 bits branching to the right. The strings in the above diagram, $0v0$ and $0v1$, are halting inputs to $U'$ with $U(0v0)\neq \perp$ and $U(0v1)\neq \perp$. Therefore, $0v$ is a total string. The infinite border sequence $\mathcal{B}\in\IS$ represents the unique infinite sequence such that all its finite prefixes have total and nontotal extensions. All finite strings branching to the right of $\mathcal{B}$ will cause $U'$ to diverge.}
		\label{fig:DomainUPrime}
	\end{center}
\end{figure}

We call this infinite sequence $\mathcal{B}$, ``border'' because for any string $x\in\FS$, $x\lhd\mathcal{B}$ implies that $x$ is total with respect to $U$ and $\mathcal{B}\lhd x$ implies that $U$ will never halt when given $x$ as an initial input. Figure~\ref{fig:DomainUPrime} shows the domain of $U'$ with respect to $\mathcal{B}$. We now set $U$ to be equal $U'$. Without loss of generality, as shown in Proposition \ref{prp:kequiv}, the complexity terms of this paper are defined with respect to the universal left-total machine $U$.

For total string $b$, we define the busy beaver function, $\bb(b)=\max\{\|x\|:U(p)=x, p\lhd b\textrm{ or }p\sqsupseteq b\}$. For total string $b$, the $b$-computable complexity of string $x$ with respect to string $y\in\FIS$ is $\K_b(x/y)=\min\{\|p\|:U_y(p)=x\textrm{ in $\bb(b)$ time and}\|y\|\leq\bb(b)\}$. If $b$ and $c$ are total, and $b\lhd c$, then $\K_b\geq \K_c$. In addition, if $b$ and $b^-$ are total, then $\K_b\geq \K_{b^-}$.

The following lemma shows that if a prefix of the border sequence is simple relative to a string $x$, then it will be the common information between $x$ and the halting sequence $\ch$.

\begin{lmm}[\citep{Epstein21}]
	\label{lmm:totalString}
	If $b\in\FS$ is total and $b^-$ is not, and $x\in\FS$, \\
	then $\K(b)+\I(x;\mathcal{H}/b)\lel \I(x\,{;}\,\mathcal{H})+\K(b/\langle x,\|b\|\rangle)$.
\end{lmm}
\section{Minimum Conditional Complexity}
\label{sec:bunch}	
We recall that a $(k,l)$-bunch $X$ is a finite set of strings where $2^k< |X|$ and for all $x,x'\in X$, $\K(x/x')< l$. If $l\gg k$, such as the $(k,l)$-bunch consisting of two large independent random strings, then it is difficult to prove properties about it. If $l\approx 2k$, then interesting properties emerge.
\begin{thr}
	\label{thr:klset}
	For $(k,l)$-bunch $X$, $\min_{x,y\in X, x \neq y}\K(y/x)\lel \ceil{l-2k}^++\I(X;\ch)+2\K(k,l)$.
\end{thr}
\begin{prf}
	We assume that the universal Turing machine $U$ is left-total. Let $b$ be a shortest total string such that $\K_b(y/x)< l$ for all $x,y\in X$. We have
	\begin{align}
	\label{eq:bX}
	\K(b/X)\lea \K(\|b\|,l), 
	\end{align}
	as there is a program that, when enumerating total strings of length $\|b\|$ from left to right, returns the first string with the desired properties. The first total string found is $b$, as shown in Figure \ref{fig:totalb}.
		\begin{figure}               
  \centering
  \includegraphics[width=15cm]{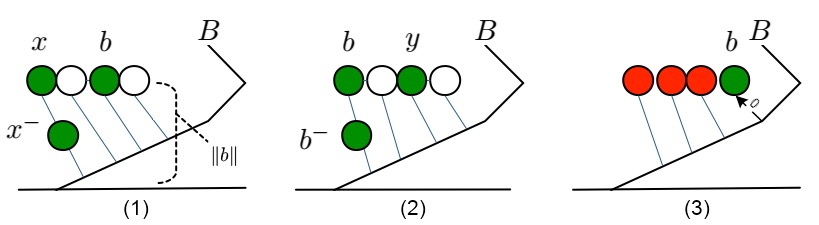}
  \caption{The above diagram represents the domain of the universal left-total Turing machine $U$ and uses the same conventions as Figure \ref{fig:DomainUPrime}, with 0s branching to the left and 1s branching to the right. It shows all the total strings of length $\|b\|$, including $b$. The large diagonal line is the border sequence, $B$. A string $c$ is marked green if $\K_c(y/x)<l$ for all $x,y \in X$. By definition, $b$ is a shortest green string. If $x$ is green and total, and $x\lhd y$, and $y$ is total, then $y$ is green, since $\K_x\geq \K_y$. Furthermore, if $x$ is green and total and $x^-$ is total, then $x^-$ is green, as $\K_x\geq \K_{x^-}$. It cannot be that there is a green $x\lhd b$ with $\|x\|=\|b\|$. Otherwise, $x^-$ is total, and thus, it is green, causing a contradiction because it is shorter than $b$. This is shown in part (1). Furthermore, there cannot be a green $y$, with $b\lhd y$ and $\|y\|=\|b\|$. Otherwise, $b^-$ is total and thus green, contradicting the definition of $b$. This is shown in part (2). Thus, $b$ is unique, and since $b^-$ is not total, by Proposition \ref{prp:unique}, $b^-$ is a prefix of the border, as shown in part (3). Thus, an algorithm returning a green string of length $\|b\|$ will return $b$.}
  \label{fig:totalb}
\end{figure}Thus, $b^-$ is not total, and by Proposition \ref{prp:unique}, $b^-\sqsubset B$ is a prefix of the border. For open parameter total string $c$, let $G^c$ be the graph defined by $(x,y)\in G_E\textrm{ iff }\K_c(y/x)< l$. Let $G=(G_E,G_V)=G^b$.  Thus if $x,y\in X$, then $(x,y)\in G_E$. We have
	\begin{align} 
	\label{eq:Gb}
	\K(G/b)&\lea \K(l)\\
	\label{eq:bG}
	\K(b/G)&\lea \K(\|b\|,l).
	\end{align}
Equation \ref{eq:Gb} is because $G=G^b$. Equation \ref{eq:bG} is due to the existence of a program that enumerates total strings of length $\|b\|$ (from left to right) and returns the first total string $c$ such that $G)E\subseteq G^c_E$. It cannot be that there is a total string $c$ shorter than $b$ with $G\subseteq G^c$. Otherwise,  $G^c_E \supseteq G_E \supseteq \binom{X}{2}$, contradicting the definition of $b$ being a shortest total string with $G^b\supseteq X$. Thus, using this impossibility and the reasoning detailed in Figure \ref{fig:totalb}, where $y$ is green if $G\subseteq G^y$, the program returns $b$. Theorem \ref{thr:graph} gives $x, y \in X$, where
	\begin{align}
	\label{eq:one}
\K(y/x) & \lel \ceil{l-2k}^++\I(X;H/G,k)+\K(G,k)
	\end{align}
	The rest of the proof is a straightforward sequence of application of inequalities. We have 
	\begin{align}
	\nonumber
	\K(X/G)&\lea \K(X/b)+\K(b/G)\\
	\label{eq:XGXb}
	&\lea \K(X/b)+\K(\|b\|,l),
	\end{align}
	where Equation \ref{eq:XGXb} is due to Equation \ref{eq:bG}.
	We also have
	 \begin{align}
	 \nonumber
	 \K(X/b,\ch)&<\K(X/G,\ch)+\K(G/b,\ch),\\
	 \label{eq:XGHl}
	 &<\K(X/G,\ch)+\K(l),
	 \end{align}
	 where Equation \ref{eq:XGHl} is due to Equation \ref{eq:Gb}. Therefore,
	 \begin{align}
	 \nonumber
	 \I(X;\mathcal{H}/G) &=\K(X/G)-\K(X/G,\mathcal{H})\\
	 \label{eq:two}
	 &\lea  \I(X;\mathcal{H}/b)+\K(l)+\K(\|b\|,l). 
	 \end{align}
	 Combining Equations \ref{eq:one} and \ref{eq:two},
	\begin{align}
	\nonumber
	\K(y/x) & \lel \ceil{l-2k}^++\I(X;\ch/b)+\K(G)+\K(\|b\|)+O(\K(k,l))\\
	\label{eq:Gtob}
	& \lel \ceil{l-2k}^++\I(X;\ch/b)+\K(b)+\K(\|b\|)+O(\K(k,l))\\
	\label{eq:Removele}
	& \lel \ceil{l-2k}^++\I(X;\ch/b)+\K(b)+O(\K(k,l)).
	\end{align}
	Equation \ref{eq:Gtob} is due to Equation \ref{eq:Gb}. Equation \ref{eq:Removele} is because the precision is $(\lel)$. Furthermore, since $b$ is total and $b^-$ is not, by Proposition \ref{prp:unique}, $b^-\sqsubset B$. The border $B$ is the binary expansion of Chaitin's Omega (see Proposition \ref{prp:unique}), so $b$ is random, with $\K(\|b\|)=O(\log \K(b))$. Using Lemma \ref{lmm:totalString} on Equation \ref{eq:Removele}, we obtain
\begin{align}
\nonumber
\K(y/x)& \lel \ceil{l-2k}^++\I(X;\ch)+\K(b/X,\|b\|)+O(\K(k,l))\\
\label{eq:final}
	 & \lel \ceil{l-2k}^++\I(X;\ch)+O(\K(k,l))
	 \end{align}
	 where Equation \ref{eq:final} is due to Equation \ref{eq:bX}. Adding $( k,l)$ to the conditional on all terms results in
	 \begin{align*}
	\K(y/x,k,l) &\lel\ceil{l-2k}^++\I(X;\ch/k,l)+O(\K(k,l/k,l))\\
	&\lel\ceil{l-2k}^++\I(X;\ch/k,l)\\
	\K(y/x) &\lel\ceil{l-2k}^++\I(X;\ch)+2\K(k,l).
\end{align*}
\qed
\end{prf}
%

\appendix
\section{Conservation Inequalities}
\label{sec:cons}
The following section presents some conservation inequalities to support the main result of this paper, which is the corollary in the introduction. The results and proofs are similar to those of \citep{Levin84}, except we use $\I(a;\ch)$ instead of $\I(x:y)=\K(x)+\K(y)-\K(x,y)$.

\begin{thr}
\label{thr:probcons}
For program $q$ that computes probability $p$ over $\N$, $\E_{a\sim p}\left[2^{\I(\langle q,a\rangle;\ch)}\right] \lem 2^{\I(q;\ch)}.$
\end{thr}
\begin{prf}
The goal is to prove $\sum_ap(a)\m(a,q/\ch)/\m(a,q) \lem \m(q/\ch)/\m(q)$. Rewriting this inequality, it suffices to prove $\sum_a\big(\m(q)p(a)/\m(a,q)\big)\big(\m(a,q/\ch)/\m(q/\ch)\big)\lem 1$. The term $\m(q)p(a)/\m(a,q)\lem 1$ because $\K(q)-\log p(a)\gea \K(a,q)$. Furthermore, it follows directly that $\sum_a \m(a,q/\ch)/\m(q/\ch)\lem 1$.\qed
\end{prf}

\begin{thr}
	\label{lmm:consH}
	For partial computable $f:\N\rightarrow\N$, for all $a\in\N$, $\I(f(a);\mathcal{H})\lea\I(a;\mathcal{H})+\K(f)$. 
\end{thr}
\begin{prf}
Observe that,
	\begin{align*}
	\I(a;\ch)&=\K(a)-\K(a/\ch)\\
	&\gea \K(a,f(a))-\K(a,f(a)/\ch)-\K(f)
	\end{align*}
The chain rule ($\K(x,y) \eqa \K(x) + \K(y/\K(x),x)$) applied twice results in
\begin{align*}
\I(a;\ch)+\K(f)&\gea \K(f(a))+\K(a/f(a),\K(f(a)))-(\K(f(a)/\ch)+\K(a/f(a),\K(f(a)/\ch),\ch))\\\
&\eqa \I(f(a);\ch)+\K(a/f(a),\K(f(a)))-\K(a/f(a),\K(f(a)/\ch),\ch)\\
&\eqa  \I(f(a);\ch)+\K(a/f(a),\K(f(a)))-\K(a/f(a),\K(f(a)),\K(f(a)/\ch),\ch)\\
&\gea \I(f(a);\ch).
\end{align*}
\qed
\end{prf}
\begin{cor}
\label{cor:prob}
For probability $p$ over $\N$, computed by program $q$, $\E_{a\sim p}[2^{\I(a;\ch)}] \lem 2^{\I(q;\ch)}.$
\end{cor}
\begin{prf}
This corollary follows from Theorems \ref{thr:probcons} and \ref{lmm:consH}.
\qed
\end{prf}
\begin{cor}
\label{cor:consinf}
For probability $p$ over $\N$, computed by program $q$,\\ $\Pr_{a\sim p}\left[\I(a;\ch)>\I(q;\ch)+m\right]\lem 2^{-m}$.
\end{cor}
\begin{prf}
This corollary follows from Corollary \ref{cor:prob} and Markov's inequality.
\qed
\end{prf}
\section{Warm-up Exercise in Stochasticity}
\label{sec:exerstoc}
The following proof demonstrates how the stochasticity term $\Ks$ can be used in mathematical arguments. The general structure of the proof parallels the proof in Theorem \ref{thr:graph}. This lemma first appeared (in a slightly different form) as Lemma 5 in \cite{EpsteinNoteOutliers21}. The lemma itself is just an exercise and is not used in the paper.
\begin{lmm}
	\label{lmm:stochasticity}
	For $D{\subseteq} \BT^n$, $|D|{=}2^s$, $\min_{x\in D}\K(x)\lel n-s+\Ks(D)+O(\K(s,n))$.
\end{lmm}
\begin{prf}

We put $(n,s)$ on an auxiliary tape to the universal Turing machine $U$. Thus, all algorithms have access to $(n,s)$, and all complexities implicitly have $(n,s)$ as conditional terms. This can be done because the precision of the lemma is $O(\K(s,n))$. Let $Q$ realize $\Ks(D)$, with $d=\max\{\d(D|Q),1\}$. Thus, $Q$ is an elementary probability measure over $\FS$ and $D\in\mathrm{Support}(Q)$, with randomness deficiency $d$.

Let $F\subseteq\BT^n$ be a random set where each element $a\in\BT^n$ is selected independently with probability $cd2^{-s}$, where $c\in\N$ is chosen later. Let $\mathcal{U}_n$ be the uniform measure over $\BT^n$. $\E[\mathcal{U}_n(F)]\leq cd2^{-s}$. Furthermore, 
\begin{align*}
&\E[Q(\{G:|G|=2^s,G\subseteq\BT^n,G\cap F=\emptyset\})]\leq \sum_GQ(G)(1-cd2^{-s})^{2^s}< e^{-cd}.
\end{align*}
Thus, by the Markov inequality, $W\subseteq\BT^n$ can be chosen such that $\mathcal{U}_n(W)\leq 2cd2^{-s}$ and $Q(\{G:|G|=2^s,G\subseteq\BT^n,G\cap W=\emptyset\}) \leq e^{1-cd}$. 
\begin{align}
\label{eq:simplew}
\K(W/Q,d,c)&=O(1).
\end{align} 
It must be that $D\cap W\neq \emptyset$. Otherwise, we obtain a contradiction with the following reasoning. Let $t:\FS\rightarrow\R_{\geq 0}$ be a $Q$-test, with $t(G)=[|G|=2^s, G\subseteq \BT^n, G\cap W=\emptyset]e^{cd-1}$, and $\sum_GQ(G)t(G)\leq 1$. Thus, $t$ gives a high score to sets $G$ that do not intersect $W$. Therefore, $t(D)=e^{cd-1}$. We have
\begin{align}
\label{eq:exertwo}
\K(D/Q,d,c) &\lea \K(D/W,Q,d,c)\\
\label{eq:exerthree}
&\lea \K(D/t,W,Q,d,c)\\
\label{eq:exerone}
 &\lea -\log Q(D)t(D)\\
\nonumber
 &\lea -\log Q(D)-(\log e)cd\\
 \nonumber
(\log e)cd & \lea -\log Q(D)-\K(D/Q) + \K(d,c)\\
\nonumber
& \lea d + \K(d,c),
\end{align}
 which is a contradiction for a large enough $c$ dependent solely on the universal Turing machine. Equation \ref{eq:exertwo} is due to Equation \ref{eq:simplew}. Equation \ref{eq:exerthree} is because the test $t$ can be computed from $(W,c,d)$. Equation \ref{eq:exerone} is due to Equation \ref{eq:testsemi}. Thus, there is an $x\in D\cap W$. Thus, since $\mathcal{U}_n(W)\leq 2cd2^{-s}$, the function $q(a) = [a\in W](2^s/cd)\mathcal{U}_n(a)$ is a semimeasure. Therefore, we have 
\begin{align*}
\K(x) &\lea -\log q(x) + \K(q)\lea n +\log d -s +\K(d)+\K(Q)\lea  n  -s +\Ks(D).
\end{align*}

 \qed
\end{prf}

\noindent\textbf{Acknowledgments.} The author thanks the anonymous referees of Theoretical Computer Science for their careful review of the paper and insightful comments.


\end{document}